# Ride Acceptance Behaviour of Ride-sourcing Drivers

**Peyman Ashkrof, Corresponding author**
*Department of Transport and Planning, Delft University of Technology, Delft, Netherlands*
*Email: P.ashkrof@tudelft.nl*

**Gonçalo Homem de Almeida Correia**
*Department of Transport and Planning, Delft University of Technology, Delft, Netherlands*

**Oded Cats**
*Department of Transport and Planning, Delft University of Technology, Delft, Netherlands*

**Bart van Arem**
*Department of Transport and Planning, Delft University of Technology, Delft, Netherlands*

## Abstract

The performance of ride-sourcing services such as Uber and Lyft is determined by the collective choices of individual drivers who are not only chauffeurs but private fleet providers. In such a context, ride-sourcing drivers are free to decide whether to accept or decline ride requests assigned by the ride-hailing platform. Drivers' ride acceptance behaviour can significantly influence system performance in terms of riders' waiting time (associated with the level of service), drivers' occupation rate and idle time (related to drivers' income), and platform revenue and reputation. Hence, it is of great importance to identify the underlying determinants of the ride acceptance behaviour of drivers. To this end, we collected a unique dataset from ride-sourcing drivers working in the United States and the Netherlands through a cross-sectional stated preference experiment designed based upon disparate information conveyed to the respondents. Using a choice modelling approach, we estimated the effects of various existing and hypothetical attributes influencing the ride acceptance choice. Employment status, experience level with the platform, and working shift are found to be the key individual-specific determinants. Part-time and beginning drivers who work on midweek days (Monday-Thursday) have a higher tendency to accept ride offers. Results also reveal that pickup time, which is the travel time between the driver's location and the rider's waiting spot, has a negative impact on ride acceptance. Moreover, the findings suggest that a guaranteed tip (i.e., the minimum amount of tip that is indicated upfront by the prospective rider, a feature that is currently not available) and an additional income due to surge pricing are valued noticeably higher than trip fare. The provided insights can be used to develop customised matching and pricing strategies to improve system efficiency.

# 1. Introduction

Recent technological innovations in the mobility sector have facilitated the emergence of new modes of transport with ride-sourcing. Offering door-to-door transport services, these two-sided ride-sourcing platforms match passengers requesting rides through a mobile app with semi-independent drivers who do not only serve as chauffeurs but also act as private fleet providers. Ride-sourcing drivers mention benefiting from a considerable degree of flexibility, freedom, and independence as the most indispensable determinants for them choosing to join the platform, in one of the most prevalent examples of the gig economy (Ashkrof et al., 2020; Hall and Krueger, 2018). Drivers can freely decide where and when to drive for the platform. These choice dimensions dynamically impact the supply-demand intensity and limit the control of the central platform over drivers. Moreover, once ride-sourcing drivers decide to drive and select their working shift and area, they receive ride requests and can choose whether to accept or decline them. Drivers' choice making has far-reaching consequences for the system performance. For instance, a delayed response due to the low acceptance rate of drivers increases the waiting time of a rider and thus yielding a lower level of service. No response to a ride request decreases rider satisfaction and may affect customer retention. In both cases, this can have a direct and indirect negative impact on drivers' earnings and the platform profit. Xu et al. (2018) report that approximately 40% of the ride-hailing requests are aborted and receive no response from drivers, which carries considerable implications for the system performance.

A successful match between demand and supply is the key objective in ride-sourcing operations to safeguard the mutual interests of the actors. The rider is transported from the specified pickup point to the desired location while drivers providing the service earn money, and the platform making the matching obtains a profit. Notwithstanding, while passengers aim to minimize the trip costs, waiting and travel time, drivers' objective is to maximize their earnings and minimize idle time. The platform itself strives mostly for profit maximization and satisfying its paying customers. Hence, the matching process is non-trivial due to the need to satisfy contradictory objectives and choices of the stakeholders. That is why various policies and matching strategies are adopted to keep the balance between agents' interests. In such a novel economy, special attention should be devoted to drivers as service suppliers who make the final decision on ride requests impacting the rider satisfaction as well as the platform reputation and revenue. Nonetheless, since the entry of the ride-sourcing business into the market, the relationship between platforms and drivers has been fragile. Judging by the worldwide strikes and lawsuits filed around the world, an increasing tension has recently been observed due to the dissatisfaction of drivers with their working conditions (Hamilton and Hernbroth, 2019). Such dissatisfaction may cause distrust (Rosenblat and Stark, 2015; Wentrup et al., 2019) that can influence drivers' choices, particularly ride acceptance behaviour. Therefore, a win-win efficient matching strategy considers the utilities and limitations of all the parties through the purposeful assignment of ride requests with the nearby drivers with the highest acceptance probability. To assess this probability, it is crucial to gain a better understanding of the supply-side behavioural dynamics under different circumstances.

Research devoted to the supply side has hitherto been primarily focused on operational dimensions such as pricing strategies (Nourinejad and Ramezani, 2019; Xue et al., 2021), relocation guidance (Zha et al., 2018), matching strategies (Chen et al., 2021; Ke et al., 2021), and estimated travel time (Wang et al., 2018). In most cases, it is assumed that the fleet is operated by either fully automated vehicles which are not currently and may not be soon in operation (SAE International, 2018) or perfectly compliant rational drivers, whereas the evidence suggests that drivers' multidimensional and autonomous decisions can significantly influence the system performance.

A growing body of literature in both journalistic and academic formats have attempted to qualitatively and quantitatively investigate the labour properties of digital on-demand mobility services. Analysing a sample of around 18,400 taxi drivers working in the United States, Wang and Smart (2020) argued that the hourly income of taxi drivers has declined since the introduction of Uber. Leng et al. (2016) concluded that monetary promotion increases drivers' acceptance rate and reduces their idle time using the 40-day trip data of 9000 ride-sourcing services collected in Beijing. Zuniga-Garcia et al. (2020) proposed a framework to measure ride-sourcing driver productivity (i.e., the profit per unit time) based on the spatial and temporal variation. They found out that the principal element in ride-sourcing driver productivity is trip distance. Based on the findings, short trips result in lower productivity even in high-demand areas. Through a nine-month qualitative study into the Uber driver working experiences, Rosenblat and Stark (2015) reported that Uber manages the labour force and gains a soft control over drivers using algorithmic labour logistics such as pricing and information dissemination strategies, which constantly interact with drivers and shape their behaviour.

Ride-sourcing platforms collect and utilize historical and real-time information of the demand and supply sides to match ride requests with available drivers. This information is processed and selectively shared with the platform drivers to keep the balance between match quality (the attractiveness of a match – for both riders and drivers) and match rate (the number of matches within a specific time interval) which can conflict (Romanyuk, 2016). Aiming for a high match rate compels drivers to accept less attractive requests which leads to low match quality. On the other hand, a low match rate increases the waiting time for passengers and thereby lowering their satisfaction and loyalty. Moreover, it reduces the occupation rate of drivers, which is affecting negatively drivers' income and may contribute to traffic congestion (Beojone and Geroliminis, 2021), as well as decreases the platform revenue and its control over the workforce. Therefore, maintaining this balance improves system efficiency and the two-sided user experience.

To find such a balance, an in-depth understanding of the behaviour of individual agents is needed. Despite the extensive literature on various aspects of the demand side, the supply-side behaviour remains so far largely unknown. Conducting a focus group study with ride-sourcing drivers working in the Netherlands, Ashkrof et al. (2020) proposed a conceptual framework that characterises the relationship between tactical (working shift selection) and operational decisions (ride acceptance and relocations strategies) of drivers and the potentially related factors. They reported the distinctive behaviour between part-time and full-time drivers, as well as beginning and experienced drivers. In a closely related paper, Xu et al. (2018), found that ride requests with economic incentives such as surge pricing are more likely to be accepted by drivers. To the best of our knowledge, our research is the first study that attempts to comprehensively investigate the quantitative effects of various existing and hypothetical determinants on drivers' ride acceptance behaviour through undertaking a cross-sectional stated preference (SP) survey. The findings can provide new insights for algorithm developers, platform providers, policymakers, and researchers working in this field. The focus of this original empirical study is on the unique data collected from Uber and Lyft drivers working in the US where the ride-sourcing platforms have emerged and thrived. Moreover, the target group is extended also to drivers working for Uber and ViaVan (a European shared on-demand transit service) in the Netherlands to tentatively examine the transferability of the results to the European context. Since the survey has been conducted during the pandemic crisis, we also examine the effects of related views and attitudes on drivers' ride acceptance choices.

The remainder of this paper is organised as follows: Section 2 explains the methodologies applied for the data collection and the data analysis processes. Section 3 focuses on the study results including the descriptive analysis, the exploratory factor analysis, and the choice modelling estimation. Lastly, the findings are discussed and the paper is concluded in Section 4.

## 2. Methodology

### 2.1. Choice Modelling

Due to the binary decision of accepting or declining a request, the choice modelling approach is applied to analyse the data at the disaggregated level and estimate the effects of the identified attributes. This method is based on the probabilistic choice theory that assumes that the decision-making process has a probabilistic nature (Bierlaire and Lurkin, 2020; Hensher et al., 2005; McFadden, 1974). Although humans are presumed to be deterministic utility maximizers, the full specifications of the utilities are unknown to the analyst. This causes stochasticity that is addressed by the so-called Random Utility Maximisation (RUM) approach capturing the unexplained variation using random variables. The utility function of alternative $j$ for individual $n$ is mathematically formulated as follows:

$$U_{jn} = V_{jn} + \varepsilon_{jn} \qquad \text{Eq. (1)}$$

Where $V_{jn}$ and $\varepsilon_{jn}$, which are typically assumed to be two independent and additive contributors of the utility function, represent the systematic (deterministic) part and the error term (random parameter), respectively. $V_{jn}$ is assumed to be a linear association of the observed variables presented in Eq. (2):

$$V_{jn} = \sum_{k=1}^{k} \beta_{jk}.x_{jk} + \sum_{m=1}^{m} \beta_{jm}.x_{jm} + \sum_{l=1}^{L} \beta_{jl}.x_{jl} \qquad \text{Eq. (2)}$$

The first term includes the instrumental variables ($x_{jk}$) that are incorporated in the SP choice sets such as drivers' spatiotemporal status, passenger characteristics, and ride attributes. The second component is associated with the individual-specific attributes ($x_{ik}$) such as socio-demographic characteristics of the drivers. The third component ($x_{ik}$) corresponds to the corona-related attitudes. $\beta_{jk}$, $\beta_{jm}$, $\beta_{jl}$ represent the marginal impacts of the instrumental attributes, individual-specific factors, and attitudinal variables respectively.

Given that the attitudes of individuals cannot be observed directly, a set of measurable variables are defined to identify the attitudinal factors and include these latent variables in the deterministic part of the utility function. The so-called Hybrid Discrete Choice (HDC) model integrates the latent and explanatory variables either sequentially or simultaneously (Ben-Akiva et al., 2002). To capture drivers' attitudes toward the Covid-19 pandemic, the latent variables were initially identified by conducting an Exploratory Factor Analysis (EFA). Thereafter, a sequential approach was used to incorporate the factor scores of the latent constructs into the systematic utility.

The second component in the utility function is the error term that captures the unobserved effects and randomness in choices. This component is constructed based on distributional assumptions on the joint distribution of the error term vector, $\varepsilon_n = (\varepsilon_{1n}, \ldots, \varepsilon_{in})$. It is typically assumed that the random variables are independently and identically distributed (IID) under an EV1 (Extreme Value Type 1) distribution: $EV(\eta, \mu)$, with $\mu > 0$.

Based on the RUM method, the probability of alternative $i$ to be chosen by individual $n$ from the binary choice set $C\{i, j\}$ is equal to the probability that the respective utility of alternative $j_1$ is larger than the utility of alternative $j_2$. Eq. (3) represents the probabilistic model:

$$P_n(i|\{j_1, j_2\}) = \Pr(U_{j_1n} \geq U_{j_2n}) = \Pr(V_{j_1n} + \varepsilon_{j_1n} \geq V_{j_2n} + \varepsilon_{j_2n})$$

$$= \Pr(\varepsilon_{j_2n} - \varepsilon_{j_1n} \leq V_{j_1n} - V_{j_2n}) \quad \text{Eq. (3)}$$

In other words, Eq. (3) reflects that the probability of choosing alternative $j_1$ by individual n is dependent on the observed attractiveness of alternative $j_1$ over alternative $j_2$ ($V_{j_1n} - V_{j_2n}$) and also the difference in random terms ($\varepsilon_{j_2n} - \varepsilon_{j_1n}$).

The software package PandasBiogeme (Bierlaire, 2020) was employed to estimate the choice models using the Maximum Likelihood Estimation (MLE) approach. The objective of MLE is to find parameter estimates by maximising the likelihood function which includes the choice probabilities related to the alternatives chosen in the data. The likelihood function is formulated in Eq. (4):

$$L_{NS} = \prod_{n=1}^{N} \prod_{s=1}^{S} \prod_{J=1}^{J} (P_{nsj})^{y_{nsj}} \quad \text{Eq. (4)}$$

Where $s$ is the number of choice tasks shown to individual $n$, , $P_{nsj}$ represents the choice probability obtained from the model, and $y_{nsj}$ is a dummy variable that is equal to 1 if alternative $j$ from the choice set $S$ is chosen by individual $n$, and 0 otherwise.

### 2.2. Choice Experiment Design

Central operators apply various information-sharing policies which yield a partial disclosure of information about ride requests and the characteristics of passengers and drivers. Such policies are adopted by ride-hailing platforms which leverage on the inherent asymmetry in access to information, providing drivers with limited information when making work-related decisions. Specifically, ride acceptance behaviour is affected by such policies that restrain the thorough assessment of the ride quality (Ashkrof et al., 2020). In both the US and the Netherlands, the information provided to drivers is remarkably limited. Most notably, trip fare and final destination are not shown to drivers before ride acceptance. This so-called blind passenger acceptance is meant to avoid destination-based discrimination (Smart et al., 2015) but at the same time, it can decrease the income for drivers (Rosenblat and Stark, 2015). Despite such ambiguity, drivers can still evaluate the attractiveness of incoming requests based on the available information to maximize the utility of ride acceptance (Ashkrof et al., 2020).

In this study, two scenarios are defined based on the platform information-sharing policy: Baseline Information Provision (BIP) and Additional Information Provision (AIP). In both scenarios, drivers are requested to decide whether to accept or decline ride requests based on a finite set of information provisioned. The BIP scenario mimics the current system operations where a driver needs to decide on the ride request based upon their current spatiotemporal status, ride attributes, and passenger characteristics. Then, some additional - currently unavailable - information such as monetary features about the same request, is added in the AIP scenario giving drivers a second chance to make a choice. This enables investigating which and to what extent factors impact the decision of drivers in the existing system setting, as well as examining drivers' response to the information that is not currently available for them. Moreover, some studies including Morshed et al. (2021) argue that the covid-19 pandemic has influenced the demand side which can potentially affect how drivers make choices such as accepting more/fewer rides, changing working shift or relocation strategies. That is why the attitudes of drivers towards the pandemic are also investigated in this research. To this end, a cross-sectional SP survey has been designed to collect the required data for further analysis.

Figure 1 illustrates the information provision set-up in the SP choice experiment. Drivers receive a ride request associated with certain characteristics and they then indicate their choice to accept or decline it. This is the BIP scenario that simulates what drivers presently experience and provides the following set of relevant information:

- Request time: The time when a ride request (ping) pops up.
- Waiting/idle time: The duration between the last drop-off and the incoming ride.
- Previous ride status: Whether the previous ride request has been declined or not.
- Pickup time: Travel time between driver's current location and rider's waiting location.
- Type of request: Private or shared rides.
- Rider rating: The average rating of the rider given by drivers.
- Surge pricing: A bonus for drivers offered by the platform when demand (locally) exceeds supply.
- Driver's location: The type of built environment where the driver is located.
- Long trip (30+ min): Whether the ride takes more than 30 minutes.

Once drivers make a decision, they are given more information, which is currently unavailable, about the same ride while the baseline information is still shown. The additional information in the AIP scenario includes:

- Trip fare: The gross amount of trip fare.
- Guaranteed tip: We hypothesize that passengers can indicate how much they are willing to tip when requesting a ride and this info can be shared with drivers when a ping pops up. As soon as the ride request is matched, the specified amount of tip is enforced in case the trip is successful.
- Congestion: The estimated delay between the pickup point and the destination caused by traffic congestion.

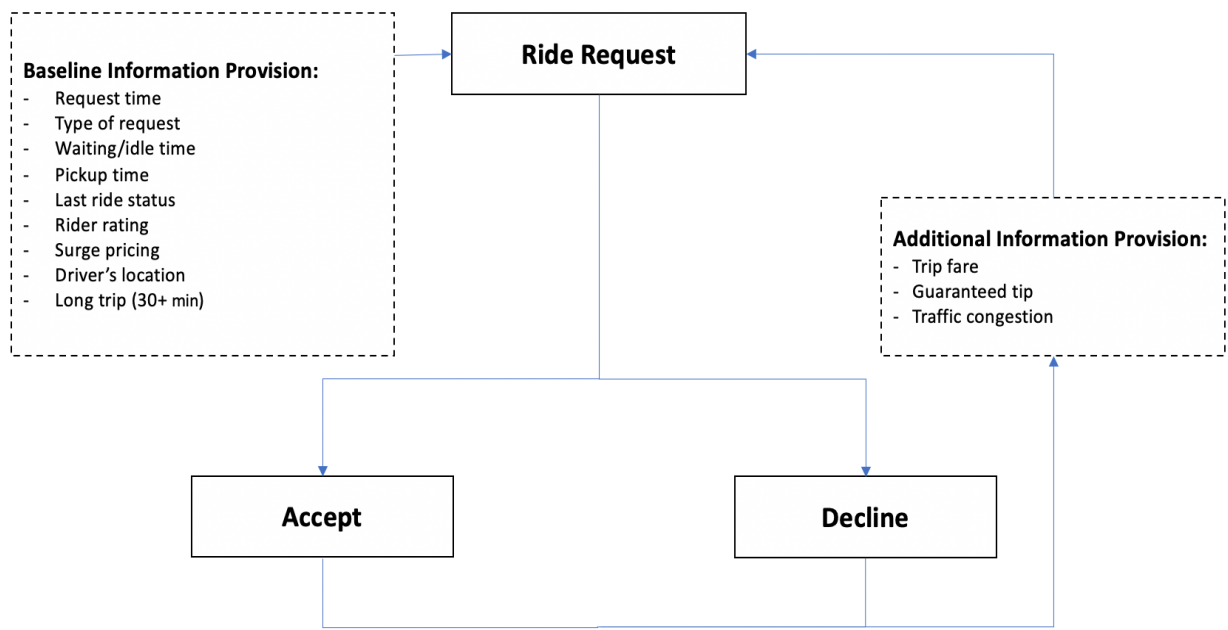

*Figure 1: Information provision set-up in the SP choice experiment*

In order to generate the experimental design of the SP survey, we identify the alternatives, attributes, and attribute levels and thereafter the type of design, model specifications, and experiment size are determined. This process is replicated with the updated input to ensure all the elements are in line with the research objectives. In the context of the choice dimension taken into account, Accept and Decline is the binary decision of drivers on ride requests which are considered as the alternatives and the information shown in each scenario are the attributes. Table 1 shows the attributes, attribute levels and labels derived by the current system operations, literature, interview with drivers, and posts on drivers' forums and then adjusted through a soft launch of the survey.

*Table 1: Attributes, attribute levels and labels*

| **BIP** | **Attributes** | **Attribute levels/labels** |
|---|---|---|
| | Request time | Pivoted around the working shift |
| | Type of request | Uber X, Uber Pool |
| | Waiting/idle time (min) | 0, 5, 10, 15 |
| | The previous request status | Declined, Accepted |
| | Rider rating (stars) | 3, 4, 5 |
| | Pick-up time (min) | 5, 10, 15, 20 |
|  | Driver's location | City centre, Suburb |
|  | Surge price | 0, 1.5, 3 |
|  | Longer than 30 min | Yes, No |
| **AIP** | Estimated trip fare | 8, 16, 24 |
|  | Guaranteed tip | 0, 1.5, 3 |
|  | Delay due to traffic congestion (min) | 0, 15, 30 |

Except for request time, the levels and labels of all the variables are specified in the table. UberX and Uber Pool refer to the private and shared-ride services, respectively. Waiting/idle time ranging from 0 to 20 minutes in this survey indicates the duration of the driver's idle status since the last drop-off. The previous request that has been declined is assumed to play a role in ride acceptance. The average rating of the riders is always shown to the drivers. Travel time between the location of driver and rider varies between 5 and 20 minutes in this experiment. The location of the driver is presumed to be either in the city centre or suburb. Surge pricing is a value that is added to the trip fare when applicable. If a trip is estimated to be taking longer than 30 minutes, drivers are notified in advance. Estimated trip fare, guaranteed tip, and the delay due to traffic congestion that are not currently available in the app are shown in the AIP scenario.

Request time is assumed to be pivoted around the reported working shift of the respondents. This is because ride-sourcing drivers can freely select their working shift and area thanks to the flexible labour model. Given that demand and supply intensity significantly varies at different times of the day as well as days of the week, drivers may have various experiences depending on the selected working shift. The pivot design ensures that drivers' can relate to the temporal characteristics of the experiment by closely resembling the experienced context to improve the response reliability. This also helps to compare the behaviour of individual drivers on different days of the week and various time slots such as peak or off-peak hours and the beginning or end of the shift.

To set up an individual-specific experiment, the segmentation procedure is applied. In this procedure, a set of designs is constructed to segment the population based upon multiple identified reference points (Rose et al., 2008). In this study, time of day is clustered into five categories: morning (5-11), midday (11-15), afternoon (15-19), evening (19-23), night (23-5) and also drivers are assumed to start their shift in one of these categories and work for either 4 hours a day (half a shift) or 8 hours a day (full shift). Therefore, the working shift in a day is divided into 10 groups as shown in table 2. Each column indicates a separate working shift that corresponds to a group of drivers. Accordingly, a library of designs is generated for the request time that has three levels in each working shift. These levels represent the beginning, the middle, and the end of the shift, respectively. Ultimately, each respondent is systematically assigned to one of these pre-defined designs based on their reported working pattern. For example, a driver who starts his/her shift at 16:00 and works for 4 hours in a day is assigned to the Afternoon 4 hours column, hence, the request time levels for this driver will be 17:00, 19:00, and 21:00.

*Table 2: Segmentation of the request time based on the working shift of drivers*

| | Morning (5-11) | | Midday (11-15) | | Afternoon (15-19) | | Evening (19-23) | | Night (23-5) | |
|---|---|---|---|---|---|---|---|---|---|---|
| | 8h | 4h | 8h | 4h | 8h | 4h | 8h | 4h | 8h | 4h |
| Request time | 8 | 8 | 13 | 13 | 17 | 17 | 21 | 21 | 2 | 2 |
| | 12 | 10 | 17 | 15 | 21 | 19 | 1 | 23 | 6 | 4 |
| | 16 | 12 | 21 | 17 | 1 | 21 | 5 | 1 | 10 | 6 |

To construct the design matrix, the efficient design method is used to generate an efficient combination of the attribute levels by minimizing the possible standard errors of the parameter estimates. These standard errors are estimated by calculating the roots of the diagonal of the asymptotic variance-covariance (AVC) matrix which is obtained from the negative inverse of the expected second derivative of the loglikelihood function of the discrete choice model as expressed in Eq. 5 and Eq. 6 (Rose and Bliemer, 2009):

$$\Omega_1 = -[E_N \left(\frac{\partial^2 logLL}{\partial \beta_{j_1 k_1} \partial \beta_{j_2 k_2}}\right)]^{-1}) \qquad \text{Eq.(5)}$$

Given:

$$\frac{\partial^2 logLL}{\partial \beta_{j_1 k_1} \partial \beta_{j_2 k_2}} = \begin{cases} \sum_{n=1}^{N} \sum_{s \in S_n} \sum_{j \in J_{ns}} x_{j_1 k_1 s} x_{j_2 k_2 s} P_{j_1 s} P_{j_2 s}, & if\ j_1 \neq j_2 \\ -\sum_{n=1}^{N} \sum_{s \in S_n} \sum_{j \in J_{ns}} x_{j_1 k_1 s} x_{j_2 k_2 s} P_{j_1 s} (1 - P_{j_2 s}), & if\ j_1 = j_2 \end{cases} \qquad \text{Eq.(6)}$$

Where $\Omega$ is the AVC matrix, $E_N$ denotes the large sample population mean, $\beta_{jk}$, $K = 1, ..., K_j$ represents the parameters of alternative $j = 1, ..., J$.

Then, the so-called D-error which is the determinant of the AVC matrix is used to set up the most efficient design with the adequately low $D - error$ (Bliemer and Rose, 2010). Since no prior information about the parameters was available, $D_z - error$ (priors equal to zero) was initially used to construct the choice sets. A pilot of 50 responses was conducted to obtain the priors. Then, $D_p - error$ was applied to minimize the standard error of the estimated parameters and reconstruct the experiment design accordingly. The following equations present the mathematical formulation of the $D - errors$:

$$D_z - error = \det(\Omega_1 (X, 0))^{1/K} \qquad \text{Eq. (7)}$$
$$D_p - error = \det(\Omega_1 (X, \beta))^{1/K} \qquad \text{Eq. (8)}$$

Where $X$ refers to the choice set design, $K$ denotes the number of parameters, and $\beta$ is the best estimate of parameters derived from the soft launch.

Moreover, two scenarios need to be designed based on the identified framework. In the BIP experiment, a set of basic information is shown to drivers and then more information is added to the existing knowledge in the AIP scenario. To implement this strategy, the model averaging method that allows multiple experiments to be evaluated at the same time is used. In this technique, the estimated AVC matrices are merged into one matrix that can be optimized for an efficiency measure such as $D-error$ (Rose and Bliemer, 2009). Therefore, both BIP and AIP models were designed simultaneously which led to a single design optimized for both models. Eventually, 24 choice sets in 4 blocks were constructed using the NGENE software package (ChoiceMetrics, 2018).

### 2.3. Questionnaire Structure

An online questionnaire instrument is used to transform the design matrix into meaningful choice sets that are randomly shown to respondents. Figure 2 displays a screenshot of the experiment interface which is carefully designed to simulate the ride request arrival process in both BIP (left) and AIP (right) scenarios.

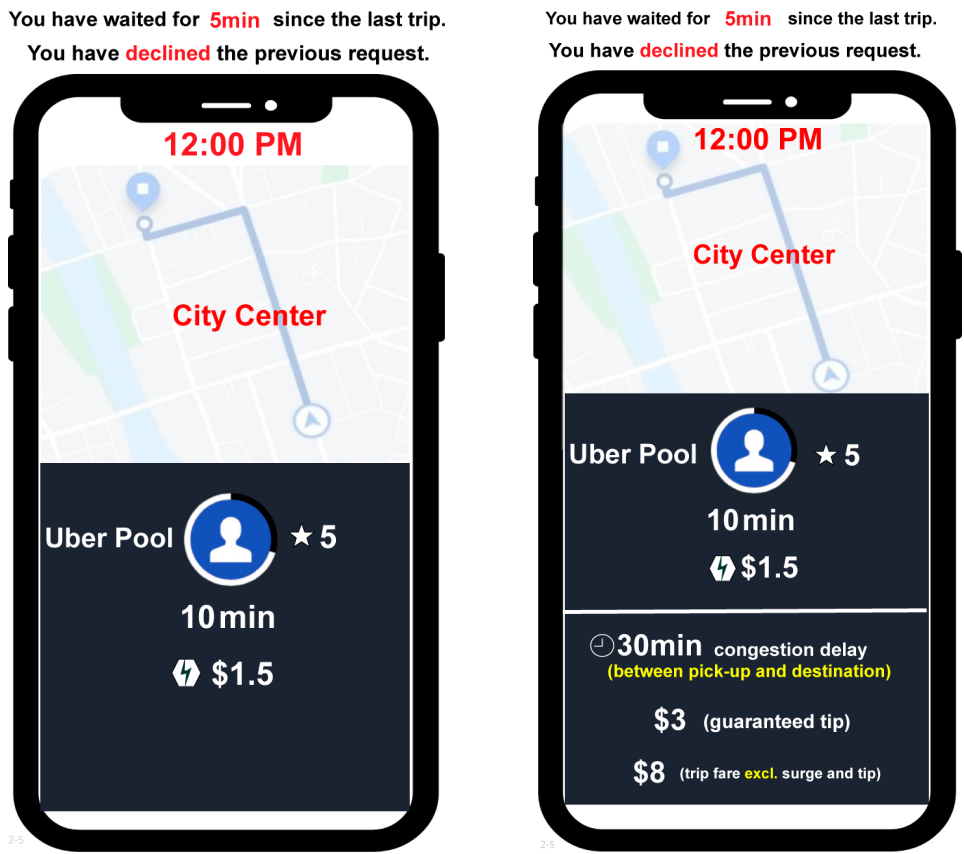

*Figure 2: Experiment interface in the BIP (left) and AIP (right) scenarios*

Furthermore, a set of screening questions is embedded at the beginning of the survey to ensure respondents are eligible to take part in this survey. The criteria are being older than 18 years old, Uber/Lyft drivers in the US or Uber/ViaVan drivers in the Netherlands, and also working at least once a week. After meeting the requirements, respondents are asked about their working pattern as input for getting assigned to the relevant design. The next part of the questionnaire is the choice experiment followed by some questions about their working pattern, employment status, experience, attitudes towards the covid-19 pandemic and their socio-economic characteristics.

## 3. Data Collection

As a highly specific target population, recruiting ride-sourcing drivers was a laborious task. A panel provider was employed to reach out to Uber and Lyft drivers in the US as well as Uber and ViaVan drivers working in the Netherlands. The data collection process took about three months from November 2020 to February 2021. Eventually, a sample of 752 and 68 drivers was drawn in the US and the Netherlands, respectively. After conducting a thorough data quality analysis, 576 responses in the US and 58 cases in the Netherlands were approved for further analysis. Despite all the efforts, a larger Dutch sample within the designated time frame was not attained due to the relatively smaller number of active ride-sourcing drivers in the Netherlands. Therefore, the focus of this study is on the US sample and the Dutch data is mainly used for a brief tentative comparative analysis.

## 4. Results

### 4.1. Descriptive Analysis

The working characteristics of the respondents are shown in Figure 3. Almost half of the drivers in the US exclusively drive for Uber while only 13% drive solely for Lyft. Multihoming strategy (i.e., working for several platforms simultaneously) is used by 41% of the respondents in the US. Uber is more dominant in the Dutch context where 77%, 2%, and 21% drive for Uber, ViaVan and both, respectively. In both countries, the majority of drivers have working experience of 13-36 months as ride-sourcing driver. Regarding the working days, Monday in the US and Saturday in the Netherlands are the most popular days to work for our sample. About 70% of the respondents start their shift in the morning and work for either 8 or 4 hours.

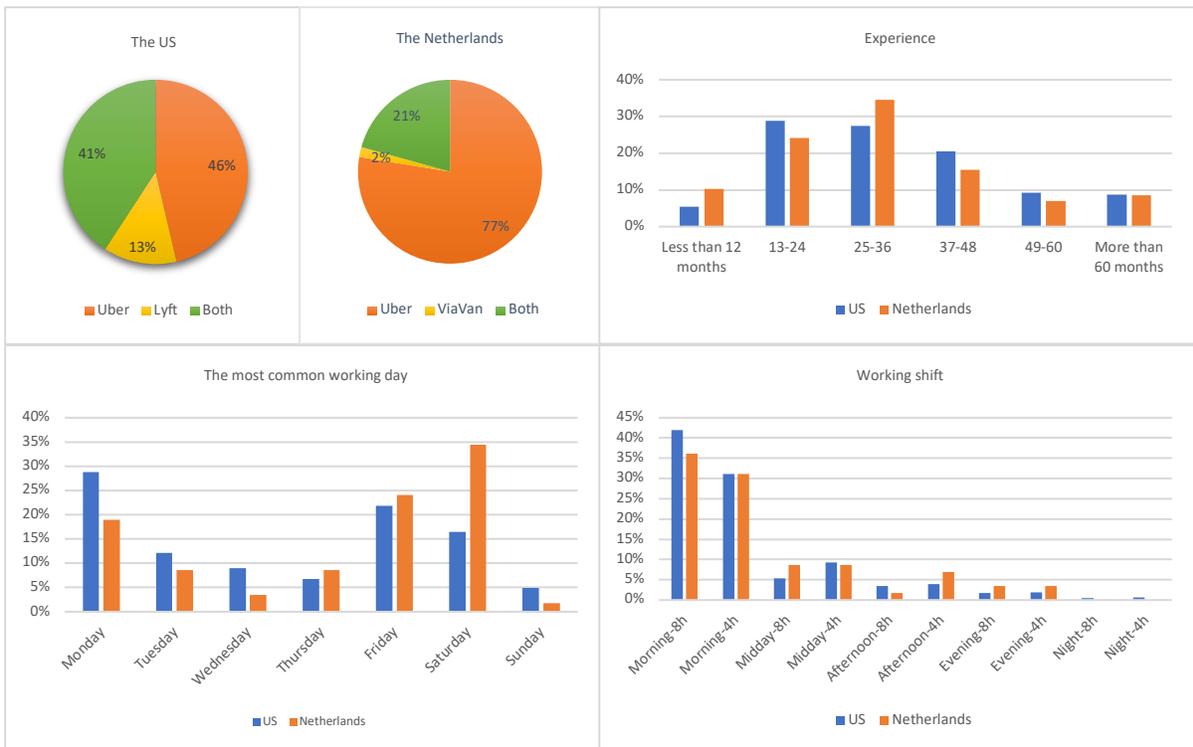

*Figure 3: Working characteristics of the drivers*

Figure 4 summarises the sociodemographic characteristics of the respondents including gender, employment status, and age. Male drivers compose more than 70% of the sample. Around 60% of the sample consists of the drivers who have other work-related sources of income, from here on labelled as part-time drivers. The data also demonstrates that the part-time drivers on average work fewer hours per week than full-time ones do. The average age of the drivers is 36 and 31 years old in the US and the Netherlands, respectively.

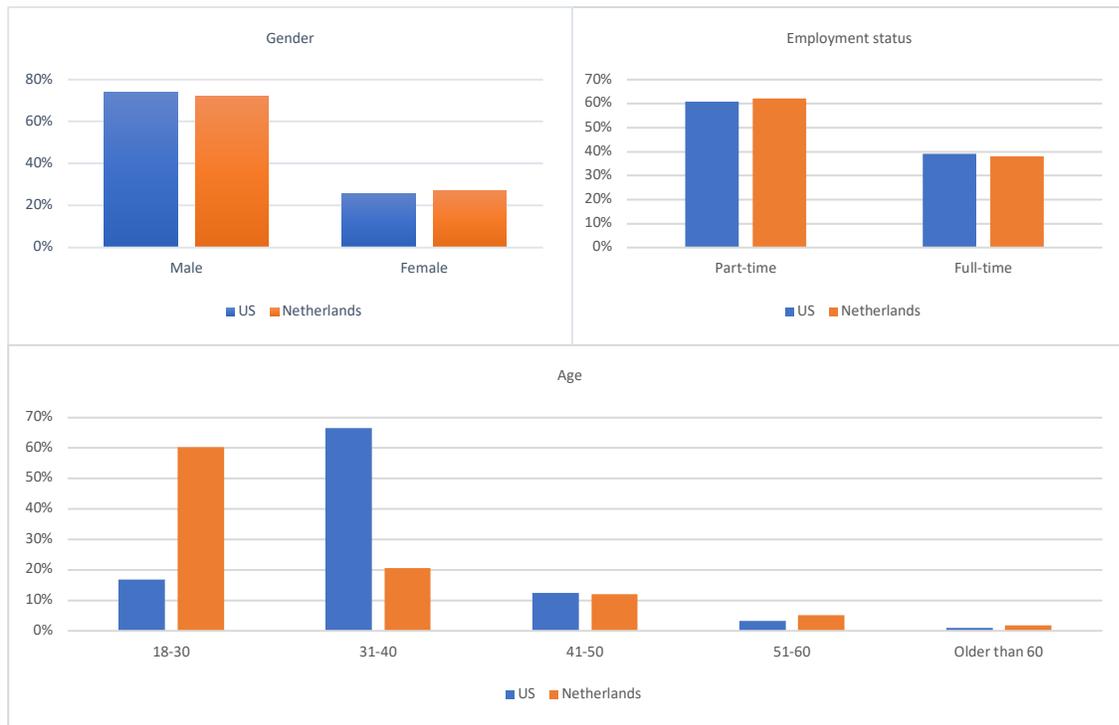

*Figure 4: Sociodemographic characteristics of the respondents*

The experience, views and attitudes of drivers towards the Covid-19 pandemic are measured by a set of statements presented in table 3. A 5-point Likert scale ranging from 1 (strongly disagree) to 5 (strongly agree) was used to capture the opinions of the respondents. The mode (the most chosen response) for each indicator is calculated to measure the central tendency of the sample in each country. Most of the drivers stated that they were concerned about the pandemic and getting infected by passengers and that they also took preventive measures to protect themselves and their clients. Furthermore, they believed that their job had been negatively affected by the pandemic. In some cases, the majority of drivers in the US and the Netherlands had different points of view. Most of the drivers working in the Netherlands neither agreed nor disagreed with changes in working shift and not driving to the busy areas while the US counterparts indicated their agreement with these statements. A contrasting viewpoint is observed between two groups of drivers about the number of incoming requests since the pandemic. The majority of the US drivers stated that they receive more requests compared to before the pandemic whereas the Dutch sampled drivers disagreed with that. Moreover, most of the drivers in the Netherlands believed that the pandemic has changed the way that they work as ride-sourcing driver while the drivers working in the US had an opposite perception.

*Table 3: The indicators measuring the attitudes of drivers towards the Covid-19 pandemic*

| No. | Statements | US Mode | NL Mode |
|---|---|---|---|
| 1 | I believe that the COVID-19 pandemic has negatively impacted my job as a driver. | 5 | 5 |
| 2 | I accept more rides than before the pandemic. | 4 | 4 |
| 3 | To comply with social distancing measures, I don't like to have more than one passenger in my car. | 4 | 3 |
| 4 | I don't care about the COVID-19. | 1 | 1 |
| 5 | I have completely changed my working shift due to the pandemic. | 4 | 3 |
| 6 | If I end up in a busy area, I don't wait there because of the risk of getting infected. | 4 | 3 |
| 7 | I'm afraid of getting infected by my passengers. | 4 | 4 |
| 8 | I don't drive to surge or high demand areas because those areas are more crowded and the risk of virus transmission is higher. | 4 | 3 |
| 9 | There is no change in what I had been doing as a driver before the pandemic. | 4 | 2 |
| 10 | I take preventive measures such as wearing a face mask, disinfecting my car, etc. to protect myself and my passengers. | 5 | 5 |
| 11 | I do care about the COVID-19. | 5 | 4 |
| 12 | I receive many more rides than before the pandemic. | 4 | 2 |

### 4.2. Exploratory Factor Analysis

To investigate the effect of the covid-19 pandemic on ride acceptance behaviour, an Exploratory Factor Analysis (EFA) was carried out to reduce the number of variables through merging the highly correlated observed measures (Henson and Roberts, 2006; Spearman, 1904). In order to ensure that the EFA is applicable, the Kaiser-Meyer-Olkin (KMO) and Bartlett's tests were performed (Kaiser, 1974). To keep a balance between parsimony and comprehensiveness, the Principal Component Analysis (PCA) model was applied (Norris and Lecavalier, 2010) and then several tests and techniques including the eigenvalues greater than 1, scree plot, and parallel analysis were deployed to ascertain the minimum number of components. Due to the superiority of the oblique rotation which takes into account the component interconnections (Flora et al., 2012; Gaskin and Happell, 2014; Price, 2017), the direct oblimin method was used to independently rotate the factor axes and situate them near the observed variables. Consequently, two components summarising the variation of the measures with the factor load greater than 0.5 were identified using the SPSS software package (Table 4).

*Table 4: Results of the exploratory factor analysis*

| Indicators | Components | |
|---|---|---|
| | 1 | 2 |
| I believe that the COVID-19 pandemic has negatively impacted my job as a driver. | | 0.659 |
| I accept more rides than before the pandemic. | 0.748 | |
| There is no change in what I had been doing as a driver before the pandemic. | 0.825 | |
| I take preventive measures such as wearing a face mask, disinfecting my car, etc. to protect myself and my passengers. | | 0.720 |
| I don't care about the COVID-19. [recoded] | | 0.696 |
| I receive many more rides than before the pandemic. | 0.848 | |
| *Extraction Method: Principal Component Analysis.* | | |
| *Rotation Method: Oblimin with Kaiser Normalization.* | | |

Given that accepting more ride requests can largely be offset by receiving many more ride offers, the first factor is mainly attributed to being positive about the pandemic effects due to having the impression of no changes (especially negative ones) in working as a driver during this period. On the other hand, the second component is primarily related to being negative about the pandemic given the stated concerns and having the perception of its negative effects on their job. Looking into the factor scores of these two components for each individual, we observed that some drivers can be associated with being either positive or negative about the pandemic and some of them have mixed feelings. Due to the relatively small Dutch sample size, the EFA was solely conducted for the US data.

### 4.3. Choice Model Estimation and Results

In total, six different models for both BIP and AIP scenarios are estimated for the US data. In each scenario, three types of models are estimated, comprising of different sets of explanatory attributes: Primary, Full, and Hybrid model. The primary model includes only the alternative-specific variables that are provided in the choice experiment. Driver's sociodemographic characteristics and working pattern are added to the ride-related attributes in the full model. The hybrid model incorporates also the experience/attitudinal factors towards the covid-19 pandemic as extracted from the EFA. This categorisation gives insights into the effects of various sets of variables depending on the application of interest. For instance, the primary model can be applied when no information about the drivers' characteristics and attitudes is available. Furthermore, the distinction between the AIP and BIP experiments is associated with the additional information shared with the drivers.

Table 5 summarises the results of the BIP model estimation including the parameter estimates, their significance value, and the model fitness. ASC_Decline represents the alternative specific constant incorporated in the utility function of the ride rejection alternative. The negative significant parameter suggests an unobserved tendency towards ride acceptance. All the other parameters are included in the utility of accepting the ride.

*Table 5: The results of the BIP models*

| Parameters | BIP | | | | | |
| --- | --- | --- | --- | --- | --- | --- |
| | **Primary** | *P-value* | **Full** | *P-value* | **Hybrid** | *P-value* |
| ASC_Decline | **-1.810** | *0.000* | **-0.417** | *0.028* | **-0.374** | *0.049* |
| B_Pickup time [min] | **-0.050** | *0.000* | - | - | - | - |
| B_Pickup time_Full [min] | - | - | **-0.027** | *0.011* | **-0.027** | *0.011* |
| B_Pickup time_Part [min] | - | - | **-0.072** | *0.000* | **-0.072** | *0.000* |
| B_Waiting time [min] | **-0.017** | *0.007* | **-0.018** | *0.005* | **-0.018** | *0.005* |
| B_Peak_hours [1=Peak hours] | **-0.560** | *0.000* | **-0.368** | *0.001* | **-0.375** | *0.000* |
| B_Weekend/Friday [1=Weekend/Friday] | **-0.443** | *0.000* | **-0.334** | *0.000* | **-0.318** | *0.000* |
| B_Time1_Loc [1=Beginning of the shift and City centre] | **-0.303** | *0.003* | **-0.284** | *0.007* | **-0.297** | *0.005* |
| B_Req_Long_Rate_Declined | **0.091** | *0.001* | **0.102** | *0.000* | **0.098** | *0.001* |
| B_Surge price [USD] | **0.101** | *0.002* | **0.110** | *0.001* | **0.108** | *0.002* |
| B_Part-time [1=Part-time drivers] | - | - | **1.110** | *0.000* | **1.120** | *0.000* |
| B_Beginners [1=Beginners] | - | - | **0.353** | *0.001* | **0.318** | *0.004* |
| B_Gender [1=Male] | - | - | **0.421** | *0.000* | **0.423** | *0.000* |
| B_Fully satisfied [1=Fully satisfied] | - | - | **0.607** | *0.000* | **0.613** | *0.000* |
| B_Education [1=Educated] | - | - | **0.080** | *0.332* | **0.135** | *0.109* |
| B_Covid_Positive | - | - | - | - | **-0.047** | *0.255* |
| B_Covid_Negative | - | - | - | - | **0.141** | *0.000* |
| **Initial Log-Likelihood** | -2395.517 | | -2395.517 | | -2395.517 | |
| **Final Log-Likelihood** | -2031.504 | | -1959.983 | | -1952.805 | |
| **Rho-square** | 0.152 | | 0.182 | | 0.185 | |

As expected, B_Pickup time which refers to the drive time from the driver's current location to the pickup point has a negative effect on ride acceptance. This is due to the fact that the pickup time increases the ride disutility since drivers are not paid while driving without a passenger. Moreover, given that no information about the trip fare and the ride destination is available in this scenario, it is not guaranteed that the incurred cost is compensated by the ride. In the full model, an interaction between the pickup time and the employment status of drivers is found significant. Part-time drivers who have other sources of income are noticeably more sensitive (almost three times) to the pickup time than full-time drivers who are entirely financially reliant on the job. This observed reluctance to take a risk may presumably stem from the more constrained working shift which makes them more conscious of time.

Another temporal component is waiting time which has a marginal negative effect on ride acceptance. Drivers' expectations may rise in relation to the time between the last drop-off and the incoming request. This is because waiting for a request leads to being idle which decreases the occupation rate and increases drivers' costs that need to be compensated. Consequently, this result suggests that drivers might prefer cherry-picking with increased elapsed waiting time.

The drivers working during the evening peak hours (16:00-00:00), weekends and Fridays, when demand is relatively higher, are more prone to decline ride requests, everything else being equal. When the frequency of incoming requests rises, drivers become more selective given that a strategical wait may lead to receiving a more profitable ride. Similarly, there is a tendency towards ride rejection at the beginning of the shift and in the city centre. These effects may be due to the expectation of having more opportunities during the remainder of the shift.

B_Req_Long_Rate_Declined suggest that there exists an interaction between request type (Uber X/Pool), long-distance trips (+30 min), rider rating, and the previously declined ride. The positive sign implies that the chance of ride acceptance is higher when a private ride (e.g. Uber X) taking more than 30 minutes is requested by a high-rated passenger while the previous request has been declined. All these components indicate a favourable ride type, one that is perceived to be profitable (long ride), less complicated (private ride), trustworthy (high-rated rider), and pressure reliever (offered after a declined ride).

As can be expected, surge pricing - a spatial-temporal pricing strategy that aims at managing supply-demand intensity - increases the probability of ride acceptance. When a request is subject to surge pricing, drivers can earn more money which incentivises them to accept it. Surge pricing which is the only monetary variable in the BIP experiment can be used to calculate the value of pickup time by computing the ratio between B_Pickup time and B_surge. Based on the results of the primary model, the value of pickup time is 0.50 USD/min. This implies that a minute increase in the pickup time can be offset by an increase of 0.50 USD in the value of surge pricing. According to the full model, this value for the full-time drivers and part-time drivers is 0.25 USD/min and 0.65 USD/min, respectively.

Among the socioeconomic factors, employment status, satisfaction degree, gender, and experience level have the highest impact on the ride acceptance behaviour, in descending order. Part-time drivers are more likely than full-time drivers to accept ride requests. This may be because they consider this job as an extra income and also their available time is limited. The level of experience also plays an important role in accepting/declining ride requests. Beginners – drivers with one year or less experience – accept rides more often. As drivers learn about the system operational strategies over time, they are better positioned to make more informed decisions. Male drivers as well as highly satisfied drivers – drivers who rated the system operations with 4.5/5 stars - have a preference for accepting rides when limited information is provided. In such a blind decision-making scenario, they may have a higher tendency to trust the platform matching algorithm.

The hybrid model that includes the corona-related factors suggests that drivers who are concerned about the pandemic and its negative effects on their work experience may have a higher acceptance rate. This group of drivers, who are prepared and protect their health by adopting preventive measures, feel the need to protect their business as well since they have the impression of the negative impact of the pandemic on their job. That is why these drivers might be willing to seize every single opportunity to earn money and compensate for those negative effects. This may lead to having a higher acceptance rate which can conflict with the match quality and the driver's income. These consequences can be the underlying reasons for the negative impression of these concerned drivers about the pandemic and its impacts. The other parameter, being positive about the pandemic, is not statistically significant in the BIP model in which the information is more restrictively shared.

Table 6 presents the results of the AIP scenario in which more information is provided to drivers. The results show that some alternative-specific factors such as waiting time, and driver's location, as well as individual-specific attributes such as working during peak hour, time of day, and gender are no longer significant. In contrast, several new alternative-specific factors including trip fare, guaranteed tip and congestion level, as well as the individual-attribute education play an important role in explaining drivers' choices. Such changes possibly stem from the importance of monetary information related to all other attributes. As expected, trip fare and tip have a positive impact on ride acceptance whereas the level of congestion indicating the delay between the pickup point and the destination motivates drivers to decline ride requests.

As observed in the BIP models, pickup time increases the disutility of accepting a ride. It should be noted that the pickup time is more negatively valued compared to the delay associated with traffic congestion. This is arguably because drivers are paid based on trip distance and travel time, so traffic congestion is possibly taken into account although not a desired experience. Driver's employment status still has significant interaction with pickup time. Part-time drivers are more sensitive to pickup time due to more constrained working hours. Additionally, the probability of accepting a ride by a part-time driver is substantially higher than for a full-time driver. Like in the BIP scenario, the interaction between request type, long ride, rider rating, and the previous declined ride is still present and leads to higher ride acceptance.

Drivers' ride acceptance behaviour can be greatly affected if ride-sourcing platforms ask riders in advance about their minimum willingness to tip and then share this information with drivers when the request appears. Once the request is accepted by the driver, the specified amount of tip is automatically secured if the driver successfully picks up the rider. The results of the primary model suggest that drivers are roughly two times more sensitive to tip and surge price than to trip fare. In other words, one monetary unit of tip and surge is worth at least two monetary units of trip fare. This effect stems from tip and surge being considered as an add-on to drivers' income. Moreover, no platform service fee is deducted from the tip while trip fare and surge pricing are subject to the commission fee (which can be about 25%). It also turns out that there is a statistically significant effect for the interaction between the guaranteed tip and the employment status of drivers. Full-time drivers are more responsive to tip than their part-time counterparts.

In this experiment, the sensitivity to the pickup time and traffic congestion can be benchmarked against the three monetary variables. The values of pickup time based on the trip fare, surge pricing, and the guaranteed tip are 1.36 USD/min, 0.71 USD/min, and 0.59 USD/min, respectively. The trade-offs for the delay time due to traffic congestion are 0.28 USD/min, 0.15 USD/min, 0.12 USD/min respectively. This suggests that monetary promotions are relatively cheaper pricing strategies than the trip fare to compensate for the pickup time as well as the delay caused by a traffic jam.

*Table 6: The results of the AIP models*

| Parameters | AIP | | | | | |
|---|---|---|---|---|---|---|
| | **Primary** | *P-value* | **Full** | *P-value* | **Hybrid** | *P-value* |
| ASC_Decline | **-1.560** | *0.000* | **-0.388** | *0.116* | -0.321 | *0.191* |
| B_Pickup time [min] | **-0.053** | *0.000* | - | - | - | - |
| B_Pickup time_Full time [min] | - | - | **-0.021** | *0.092* | **-0.021** | *0.108* |
| B_Pickup time_Part time [min] | - | - | **-0.076** | *0.000* | **-0.076** | *0.000* |
| B_Waiting time [min] | **-0.005** | *0.522* | **-0.005** | *0.518* | **-0.004** | *0.583* |
| B_Peak_hours [1=Peak hours] | **-0.057** | *0.629* | **0.027** | *0.825* | **0.022** | *0.860* |
| B_Weekend/Friday [1=Weekend/Friday] | **-0.507** | *0.000* | **-0.412** | *0.000* | **-0.412** | *0.000* |
| B_Time1_Loc [1=Beginning of the shift and City centre] | **-0.135** | *0.252* | **-0.137** | *0.253* | **-0.155** | *0.195* |
| B_Req_Long_Rate_Dec | **0.086** | *0.011* | **0.087** | *0.011* | **0.081** | *0.017* |
| B_Surge price [USD] | **0.075** | *0.048* | **0.076** | *0.049* | **0.069** | *0.074* |
| B_Trip Fare [USD] | **0.039** | *0.000* | **0.041** | *0.000* | **0.040** | *0.000* |
| B_Guaranteed tip [USD] | **0.090** | *0.014* | - | - | - | - |
| B_Guaranteed tip_Full time [USD] | - | - | **0.208** | *0.000* | **0.211** | *0.000* |
| B_Guaranteed tip_Part time [USD] | - | - | **0.021** | *0.647* | **0.015** | *0.743* |
| B_Traffic congestion [min] | **-0.011** | *0.002* | **-0.011** | *0.002* | **-0.012** | *0.001* |
| B_Part-time [1=Part-time drivers] | - | - | **0.981** | *0.000* | **1.03** | *0.000* |
| B_Beginners [1=Beginners] | - | - | **0.271** | *0.023* | **0.223** | *0.062* |
| B_Gender [1=Male] | - | - | **0.113** | *0.259* | **0.124** | *0.215* |
| B_Fully satisfied [1=Fully satisfied] | - | - | **0.190** | *0.029* | **0.217** | *0.012* |
| B_Education [1=Educated] | - | - | **0.461** | *0.000* | **0.561** | *0.000* |
| B_Covid_Positive | - | - | - | - | **-0.121** | *0.010* |
| B_Covid_Negative | - | - | - | - | **0.178** | *0.000* |
| **Initial Log-Likelihood** | -2395.517 | | -2395.517 | | -2395.517 | |
| **Final Log-Likelihood** | -1752.026 | | -1722.981 | | -1710.417 | |
| **Rho-square** | 0.269 | | 0.281 | | 0.286 | |

Although the education level was not found to be an influential factor in the restricted information-sharing policy, the results of the AIP models indicate that drivers that attained higher levels of education (i.e. have a college or a higher degree) are more likely to accept rides. Like in the BIP experiment, beginning and fully satisfied drivers tend to accept more rides. Beginning drivers may lack sufficient knowledge of the system operations to evaluate the ride quality and fully satisfied drivers have a higher trust in the system performance.

Regarding the coronavirus pandemic effects, the two identified factors are incorporated into the AIP model. Unlike the results of the BIP hybrid model, being positive about the pandemic is statistically significant and has a negative effect on ride acceptance. This component is obtained from three attitudinal statements about accepting more rides that can be offset with receiving many more ride requests than before the pandemic, and having the perception of no changes in work before and during the pandemic. These drivers have the impression of receiving notably more requests. Although the evidence shows that the total number of requests has declined (Du and Rakha, 2021), some drivers have stopped working given the more dramatic plunge in demand at the beginning of the pandemic as well as the high risk of getting infected. This may have decreased the competition between some groups of drivers and increased their chance to receive ride requests Therefore, receiving more requests or at least having such an impression makes drivers more selective and causes more rejection. Conversely, being negative about the pandemic can increase the chance of acceptance as observed in the BIP scenario. In the AIP scenario, these two components have opposite values that can offset each other. Therefore, the pandemic may not significantly influence the ride acceptance behaviour of drivers at the aggregate level of this scenario.

Due to the relatively small dataset collected in the Netherlands, we could not estimate a statistically sound separate model for the Dutch sample. Alternatively, the data from both countries were merged after unifying the attribute units, allowing the analysis of the combined sample and identifying the possible differences in drivers' behaviour by specifying dummy variables. Among the estimated models, the following differences between the two groups of drivers in the AIP-Primary model were found. Sensitivity to traffic congestion was much higher among the drivers working in the Netherlands, possibly because the level of congestion is lower in the Netherlands, according to the traffic index (Traffic Index by Country, 2021). Furthermore, the trip fare was regarded nearly two times more important in the Netherlands than in the US. There may exist multiple underlying reasons including the currency, tipping culture (which is less customary in the Netherlands than in the US), income level, and other economic indices. However, these observations need to be further investigated with a larger sample size in the Netherlands in order to draw more conclusive results.

## 5. Discussion and Conclusions

This research unravels the ride acceptance behaviour of ride-sourcing drivers through a stated preference experiment performed in the United States and the Netherlands. To the best of our knowledge, this is the first study attempting to comprehensively estimate the determinants of ride-sourcing drivers' ride acceptance behaviour. To this end, a set of potential attributes are identified based on the current system operations, driver-side app, existing literature, interview with drivers, and posts on drivers' forums. Then, two information-sharing policies are defined: Baseline Information Provision (BIP) and Additional Information Provision (AIP). The former scenario solely includes the variables currently shown to drivers in the most commonly used system setting while additional information is provided in the subsequent phase of the experiment. In total, 576 and 56 qualified responses were collected in the US and the Netherlands, respectively. Subsequently, a choice modelling approach is applied to analyse the data. The focus of this study is on the US data due to the relatively small sample size in the Netherlands.

The monetary variables included in this study are surge pricing, trip fare, and guaranteed tip (i.e., the minimum amount of tip that is indicated upfront by the prospective rider). Surge pricing included in the BIP experiment is the only monetary attribute that is shared with drivers in the current system setting of the ride-sourcing platforms operating in the target area whereas trip fare and guaranteed tip are incorporated in the AIP scenario. Results reveal that guaranteed tip is the most highly valued monetary factor, especially for full-time drivers who are more financially dependent on the ride-sourcing platforms, followed closely by surge pricing. From the drivers' perspective, tip and surge pricing as added income are considered about two times more important than trip fare.

In general, tipping is a pro-social consumer behaviour that is considered as an economically irrational action of customers and typically targets the low-income service providers (Azar, 2003; Elliott et al., 2018). Such a social norm has a profound economic impact on the US service industry (Shierholdz et al., 2017). In the US taxi industry in 2012, tipping comprised around 18% of the annual taxi revenue which is equal to $445 million (Bloomberg and Yassky, 2014). Currently, Uber riders can tip after they are dropped off. Analysing 40 million observations of Uber tipping behaviour in 2017, Chandar et al. (2019) concluded that more than 15% of the trips are tipped although tips are given privately (no consequences for rider rating) and the chance of having a match with the same driver is fairly low. They also found out that the average amount of tip is approximately $0.5 per trip and for those rides that have been tipped, more than $3 is tipped which is about 26% of the trip fare. In this study, we have introduced a new form of tipping that is determined in advance. When the ride is matched, the specified amount of tip must be paid and naturally, the passenger can tip more to reward the service if satisfying.

This feature can be used when a rider highly disvalues waiting time (e.g., being in a hurry) and intends to persuade nearby drivers to quickly accept the ride. It is effectively a self-determined discriminatory pricing scheme that allows riders to signal their willingness to pay and thereby potentially influencing the level of service received. This is in line with the study conducted by Flath (2012) which suggests that passengers with a strong aversion to waiting would tip taxi drivers to reduce the time needed to find a taxi. As opposed to trip fare and surge pricing, tipping is not directly imposed on riders by the platform which makes it less unfavourable from the rider's perspective. The results of this study suggest that such a feature can significantly impact drivers' ride acceptance behaviour. This can also be part of the platform pricing strategy through developing an algorithm that optimally calculates the trip fare and surge pricing based on the guaranteed tip determined by riders. This may lead to a higher acceptance rate and level of service which is beneficial for riders, drivers, and the platform.

Surge pricing is a spatial-temporal pricing strategy that is introduced to address an imbalanced supply-demand relation. However, surge pricing is one of the most controversial topics in the ride-sourcing literature given its enormous implications for all stakeholders involved. On one hand, it is argued that surge pricing is a near-optimal solution that decreases the match failure as well as system inefficiency through suppressing the excessive demand and also increases the platform profit (Cachon et al., 2017; Nourinejad and Ramezani, 2019). Using machine learning techniques, Battifarano et al. (2019) propose that surge pricing can generate more profit if the value is predicted and disseminated to both riders and drivers in advance. On the other hand, surge pricing may lead to strategic waiting for both riders who seek normal price and drivers looking for higher prices which results in inefficient performance due to forward-looking behaviour (Ashkrof et al., 2020; Chen and Hu, 2020; Zhong et al., 2020). The results of this study indicate that surge pricing is an important determinant of ride acceptance behaviour by ride-sourcing drivers. This is in line with the findings of Chen et al. (2015). They found that drivers work longer and flexibly adjust their working shift when surge pricing is present even if they have already hit their daily target. Based on the findings of this research, surge pricing is the second most important monetary attribute that can strongly incentivise drivers to accept rides. The value of pickup time for surge pricing is estimated to be 0.5-0.71 USD/min. This has important consequences for determining the expected response of drivers to the introduction of surge pricing as a function of their travel time from the surge location and the surge price level. Unlike the guaranteed tip, no difference in perspectives of part-time and full-time drivers concerning surge pricing is found.

Nevertheless, employment status is a crucial attribute influencing the choice of drivers. Part-time drivers, who have other sources of income, show a strong preference for accepting ride offers compared to their full-time counterparts. This might be because part-time drivers supplement their revenue from other jobs and also have limited available time restricting their degrees of freedom. Hence, the opportunity costs of part-time drivers are potentially higher which leads to a higher acceptance rate (Baron, 2018).

Furthermore, the experience level of drivers with the ride-sourcing platforms and their operational strategies has been identified as a determinant that influences their choices in various aspects (Chu et al., 2018; Miranda et al., 2008; Noulas et al., 2019; Rosenblat and Stark, 2015; Wang and Yang, 2019). Based on the findings of this study, beginning drivers who have one year or less of experience with ride-hailing tend to accept more rides. Lack of sufficient experience and knowledge to evaluate the characteristics of ride requests and having higher trust in the system performance might be the underlying reasons for this tendency (Ashkrof et al., 2020). In both BIP and AIP experiments, pickup time, especially for part-time drivers, has a negative impact on ride acceptance due to the disutility of driving without a passenger, i.e. unpaid time. Therefore, in order to have a higher acceptance rate, a new matching algorithm can be developed that can calculate the response likelihood of nearby drivers and then offer the request to the driver with the highest probability of acceptance. For instance, less attractive requests can be matched with part-time beginning drivers. The introduction of such measures should consider their potential acceptance amongst drivers.

While the small sample collected in the Netherlands, does not allow for estimating a full-fledged model, it has been observed that drivers working in the Netherlands are more sensitive to the trip fare as well as traffic congestion. These findings should be further investigated with a larger sample size from the Netherlands and possibly from other European countries. Another limitation of this research refers to the inherently typical bias of stated preference surveys in which respondents may not accurately grasp the choice experiments, especially the AIP scenario that includes several hypothetical new components. It can be insightful to validate the findings of this study through analysing a set of revealed preferences data concerning drivers' behaviour in ride-sourcing environments or field observation of drivers if possible. Moreover, the insights gained in this study can be integrated into ride-hailing analysis models (Kucharski and Cats, 2020) and used to assess the possible effects of driver's ride acceptance behaviour based on various information-sharing policies on the ride-sourcing system performance, including efficiency, level-of-service and profitability. Future research may investigate other aspects of ride-sourcing drivers' decisions such as registration to the platform at the strategic level, selecting working shift at the tactical level, and relocation strategies at the operational level.

## Acknowledgements

This research was supported by the CriticalMaaS project (grant number 804469), which is financed by the European Research Council and the Amsterdam Institute for Advanced Metropolitan Solutions.

## References

Ashkrof, P., Correia, G.H. de A., Cats, O., van Arem, B., 2020. Understanding ride-sourcing drivers' behaviour and preferences: Insights from focus groups analysis. Res. Transp. Bus. Manag. 37, 100516. https://doi.org/10.1016/j.rtbm.2020.100516

Azar, O.H., 2003. The implications of tipping for economics and management. Int. J. Soc. Econ. 30, 1084–1094. https://doi.org/10.1108/03068290310492878

Baron, D.P., 2018. Disruptive Entrepreneurship and Dual Purpose Strategies: The Case of Uber. Strateg. Sci. 3, 439–462. https://doi.org/10.1287/stsc.2018.0059

Battifarano, M., Qian, Z. (Sean), 2019. Predicting real-time surge pricing of ride-sourcing companies. Transp. Res. Part C Emerg. Technol. 107, 444–462. https://doi.org/10.1016/j.trc.2019.08.019

Ben-Akiva, M., Walker, J., Bernardino, A.T., Gopinath, D.A., Morikawa, T., Polydoropoulou, A., 2002. Integration of Choice and Latent Variable Models, In Perpetual Motion. https://doi.org/10.1016/b978-008044044-6/50022-x

Beojone, C.V., Geroliminis, N., 2021. On the inefficiency of ride-sourcing services towards urban congestion. Transp. Res. Part C Emerg. Technol. 124, 102890. https://doi.org/10.1016/j.trc.2020.102890

Bierlaire, M., 2020. A Short Introduction to PandasBiogeme. Ec. Polytech. Fed. Lausanne 13–43.

Bierlaire, M., Lurkin, V., 2020. Introduction to choice models [WWW Document]. URL https://www.edx.org/course/introduction-to-discrete-choice-models

Bliemer, M.C.J., Rose, J.M., 2010. Construction of experimental designs for mixed logit models allowing for correlation across choice observations. Transp. Res. Part B Methodol. https://doi.org/10.1016/j.trb.2009.12.004

Bloomberg, M.R., Yassky, D., 2014. 2014 Taxicab Factbook. Taxicab Factb. 1–13.

Cachon, G.P., Daniels, K.M., Lobel, R., 2017. The role of surge pricing on a service platform with self-scheduling capacity. Manuf. Serv. Oper. Manag. 19, 368–384. https://doi.org/10.1287/msom.2017.0618

Chandar, B., Gneezy, U., List, J.A., Muir, I., 2019. The Drivers of Social Preferences : Evidence from a Nationwide Tipping Field Experiment.

Chen, M.K., Sheldon, M., 2015. Dynamic Pricing in a Labor Market : Surge Pricing and Flexible Work on the Uber. Ssrn 1–19.

Chen, Xiqun, Chen, Xiaowei, Zheng, H., Xiao, F., 2021. Efficient dispatching for on-demand ride services : Systematic optimization via Monte-Carlo tree search. Transp. Res. Part C 127, 103156. https://doi.org/10.1016/j.trc.2021.103156

Chen, Y., Hu, M., 2020. Pricing and matching with forward-looking buyers and sellers. Manuf. Serv. Oper. Manag. 22, 717–734. https://doi.org/10.1287/msom.2018.0769

ChoiceMetrics, 2018. Ngene 1.2 USER MANUAL & REFERENCE GUIDE The Cutting Edge in Experimental Design End-User License Agreement 241.

Chu, L.Y., Wan, Z., Zhan, D., 2018. Harnessing the Double-Edged Sword via Routing: Information Provision on Ride-Hailing Platforms. SSRN Electron. J. https://doi.org/10.2139/ssrn.3266250

Du, J., Rakha, H., 2021. Impact of COVID-19 on Ridehailing and Other Modes of Transportation [WWW Document]. URL https://www.morgan.edu/school_of_engineering/research_centers/urban_mobility_and_equity_center/research/new_research/covid_and_ridehailing.html (accessed 5.13.21).

Elliott, D., Tomasini, M., Oliveira, M., Menezes, R., 2018. Tippers and stiffers: An analysis of tipping behavior in taxi trips. 2017 IEEE SmartWorld Ubiquitous Intell. Comput. Adv. Trust. Comput. Scalable Comput. Commun. Cloud Big Data Comput. Internet People Smart City Innov. SmartWorld/SCALCOM/UIC/ATC/CBDCom/IOP/SCI 2017 - 1–8. https://doi.org/10.1109/UIC-ATC.2017.8397523

Flath, D., 2012. Why Do We Tip Taxicab Drivers? Japanese Econ. 39, 69–76. https://doi.org/10.2753/jes1097-203x390304

Flora, D.B., LaBrish, C., Chalmers, R.P., 2012. Old and new ideas for data screening and assumption testing for exploratory and confirmatory factor analysis. Front. Psychol. 3, 1–21. https://doi.org/10.3389/fpsyg.2012.00055

Gaskin, C.J., Happell, B., 2014. On exploratory factor analysis: A review of recent evidence, an assessment of current practice, and recommendations for future use. Int. J. Nurs. Stud. 51, 511–521. https://doi.org/10.1016/j.ijnurstu.2013.10.005

Hall, J. V., Krueger, A.B., 2018. An Analysis of the Labor Market for Uber's Driver-Partners in the United States. ILR Rev. 71, 705–732. https://doi.org/10.1177/0019793917717222

Hamilton, I., Hernbroth, M., 2019. Uber drivers across the world are striking about pay, conditions, and the firm's "orgy of greed" [WWW Document]. businessinsider. URL https://www.businessinsider.nl/photos-uber-drivers-go-on-strike-around-the-world-2019-5?international=true&r=US (accessed 5.6.21).

Hensher, D.A., Rose, J.M., Rose, J.M., Greene, W.H., 2005. Applied Choice Analysis, Cambridge University Press.

Henson, R.K., Roberts, J.K., 2006. Use of Exploratory Factor Analysis in Published Research. Educ. Pschological Meas. 66, 11–14. https://doi.org/10.1177/0013164405282485

Kaiser, H.F., 1974. An index of factorial simplicity. Psychometrika 39, 31–36.

https://doi.org/10.1007/BF02291575

Ke, J., Zheng, Z., Yang, H., Ye, J., 2021. Data-driven analysis on matching probability, routing distance and detour distance in ride-pooling services. Transp. Res. Part C Emerg. Technol. 124, 102922. https://doi.org/10.1016/j.trc.2020.102922

Kucharski, R., Cats, O., 2020. Exact matching of attractive shared rides (ExMAS) for system-wide strategic evaluations. Transp. Res. Part B Methodol. 139, 285–310. https://doi.org/10.1016/j.trb.2020.06.006

Leng, B., Du, H., Wang, J., Li, L., Xiong, Z., 2016. Analysis of taxi drivers' behaviors within a battle between two taxi apps. IEEE Trans. Intell. Transp. Syst. 17, 296–300. https://doi.org/10.1109/TITS.2015.2461000

McFadden, D., 1974. Conditional Logit Analysis of Qualitative Choice Behaviour, in: Frontiers of Econometrics.

Miranda, F., Muñoz, J.C., De dios Ortúzar, J., 2008. Identifying transit driver preferences for work shift structures: An econometric analysis. Transp. Sci. 42, 70–86. https://doi.org/10.1287/trsc.1070.0199

Morshed, S.A., Khan, S.S., Tanvir, R.B., Nur, S., 2021. Impact of COVID-19 pandemic on ride-hailing services based on large-scale Twitter data analysis. J. Urban Manag. https://doi.org/10.1016/j.jum.2021.03.002

Norris, M., Lecavalier, L., 2010. Evaluating the use of exploratory factor analysis in developmental disability psychological research. J. Autism Dev. Disord. 40, 8–20. https://doi.org/10.1007/s10803-009-0816-2

Noulas, A., Salnikov, V., Hristova, D., Mascolo, C., Lambiotte, R., 2019. Developing and deploying a taxi price comparison mobile app in the wild: Insights and challenges. Proc. - 2018 IEEE 5th Int. Conf. Data Sci. Adv. Anal. DSAA 2018 424–430. https://doi.org/10.1109/DSAA.2018.00055

Nourinejad, M., Ramezani, M., 2019. Ride-Sourcing Modeling and Pricing in Non-Equilibrium Two-Sided Markets. Transp. Res. Part B 00, 24–26. https://doi.org/10.1016/j.trb.2019.05.019

Price, L.R., 2017. Psychometric Methods: Theory into Practice [WWW Document]. Meas. Interdisc. Res. Perspect. URL https://lccn.loc.gov/2016013346 (accessed 4.14.21).

Romanyuk, G., 2016. Ignorance Is Strength : Improving Performance of Matching Markets by Limiting Information. Job Mark. Pap. 1–43.

Rose, J.M., Bliemer, M.C.J., 2009. Incorporating model uncertainty into the generation of efficient stated choice experiments: A model averaging approach, in: International Choice Modelling Conference, March 30-April 1, Yorkshire U.K.

Rose, J.M., Bliemer, M.C.J., Hensher, D.A., Collins, A.T., 2008. Designing efficient stated choice experiments in the presence of reference alternatives. Transp. Res. Part B Methodol. 42, 395–406. https://doi.org/10.1016/j.trb.2007.09.002

Rosenblat, A., Stark, L., 2015. Algorithmic Labor and Information Asymmetries: A Case Study of Uber's Drivers. Ssrn 10, 3758–3784. https://doi.org/10.2139/ssrn.2686227

SAE International, 2018. Surface Vehicle Recommended Practice.

Shierholdz, H., Cooper, D., Wolfe, J., Zipperer, B., 2017. Employers would pocket $5.8 billion of workers' tips under Trump administration's proposed 'tip stealing' rule [WWW Document]. A Rep. by Econ. Policy Institute, Washingt. DC. URL https://www.epi.org/publication/employers-would-pocket-workers-tips-under-trump-administrations-proposed-tip-stealing-rule/ (accessed 4.29.21).

Smart, R., Rowe, B., Hawken, A., 2015. Faster and Cheaper: How Ride-Sourcing Fills a Gap in Low-Income Los Angeles Neighborhoods.

Spearman, C., 1904. " General Intelligence ," Objectively Determined and Measured Author ( s ): C . Spearman Source : The American Journal of Psychology , Vol . 15 , No . 2 ( Apr ., 1904 ), pp . 201-292 Published by : University of Illinois Press Stable URL : http://www.jsto. Am. J. Psychol. 15, 201–292.

Traffic Index by Country, 2021. Traffic Index by Country 2021 [WWW Document]. URL https://www.numbeo.com/traffic/rankings_by_country.jsp (accessed 5.13.21).

Wang, H., Yang, H., 2019. Ridesourcing systems: A framework and review. Transp. Res. Part B Methodol. 129, 122–155. https://doi.org/10.1016/j.trb.2019.07.009

Wang, S., Smart, M., 2020. The disruptive effect of ridesourcing services on for-hire vehicle drivers' income and employment. Transp. Policy 89, 13–23. https://doi.org/10.1016/j.tranpol.2020.01.016

Wang, Z., Fu, K., Ye, J., 2018. Learning to estimate the travel time. Proc. ACM SIGKDD Int. Conf. Knowl. Discov. Data Min. 858–866. https://doi.org/10.1145/3219819.3219900

Wentrup, R., Nakamura, H.R., Ström, P., 2019. Uberization in Paris – the issue of trust between a digital platform and digital workers. Crit. Perspect. Int. Bus. 15, 20–41. https://doi.org/10.1108/cpoib-03-2018-0033

Xu, K., Sun, L., Liu, J., Wang, H., 2018. An empirical investigation of taxi driver response behavior to ride-hailing requests: A spatio-temporal perspective. PLoS One 13, 1–17. https://doi.org/10.1371/journal.pone.0198605

Xue, Z., Zeng, S., Ma, C., 2021. Economic modeling and analysis of the ride-sourcing market considering labor supply. Res. Transp. Bus. Manag. 38, 100530. https://doi.org/10.1016/j.rtbm.2020.100530

Zha, L., Yin, Y., Xu, Z., 2018. Geometric matching and spatial pricing in ride-sourcing markets. Transp. Res. Part C Emerg. Technol. 92, 58–75. https://doi.org/10.1016/j.trc.2018.04.015

Zhong, Y., Wan, Z., Shen, Z.-J.M., 2020. Queueing Versus Surge Pricing Mechanism: Efficiency, Equity, and Consumer Welfare. SSRN Electron. J. https://doi.org/10.2139/ssrn.3699134

Zuniga-Garcia, N., Tec, M., Scott, J.G., Ruiz-Juri, N., Machemehl, R.B., 2020. Evaluation of ride-sourcing search frictions and driver productivity: A spatial denoising approach. Transp. Res. Part C Emerg. Technol. 110, 346–367. https://doi.org/10.1016/j.trc.2019.11.021